# Decision-level multi-method fusion of spatially scattered data from nondestructive inspection of ferromagnetic parts

René Heideklang[a], Parisa Shokouhi[b]

[a] BAM Federal Institute for Materials Research and Testing, Unter den Eichen 87, 12205 Berlin, Germany

rene.heideklang@bam.de

phone: +49 30 8104-1838 (secretariat of Division 8.5)

[b] Department of Civil and Environmental Engineering, The Pennsylvania State University, 215 Sackett Bldg., University Park, PA 16802

parisa@engr.psu.edu

phone: +1 (814) 863-0678

Corresponding author: René Heideklang

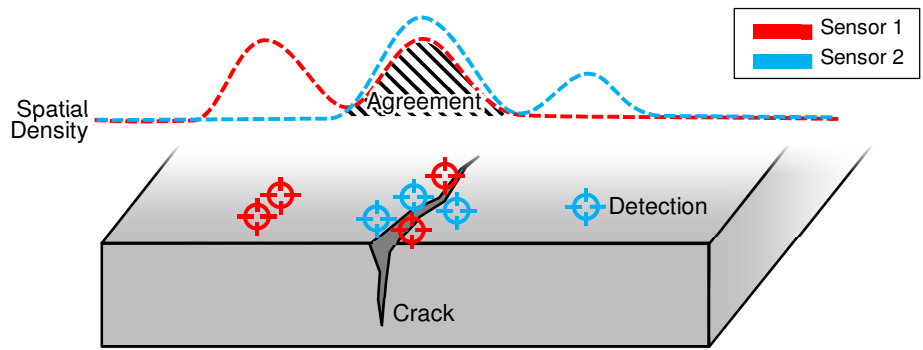

# Abstract

This article deals with the fusion of flaw detections from multi-sensor nondestructive materials testing. Because each testing method makes use of different physical effects for defect localization, a multi-method approach is promising to effectively distinguish the many false alarms from actual material defects. To this end, we propose a new fusion technique for scattered two- or three-dimensional location data. Using a density-based approach, the proposed method is able to explicitly address the localization uncertainties such as registration errors. We provide guidelines on how to set all key parameters and demonstrate the technique's robustness. Finally, we apply our fusion approach to experimental data and demonstrate its ability to find small defects by substantially reducing false alarms under conditions where no single-sensor method is adequate.

# 1 Introduction

Industrial nondestructive testing (NDT) refers to the inspection of materials, parts or structures concerning their condition without compromising their integrity. NDT experts employ different sensors depending on the material and the expected types of defects. Near-surface cracks represent one class of flaws that are commonly encountered in critical machine parts under dynamic loading such as turbine blades and bearings. Repeating cycles of dynamic stress promote the growth of invisible micro cracks into larger defects that compromise the safety and shorten the service life of the component. As such, early detection of cracking is critical to the safe operation and serviceability of flawed machinery. Among the NDT methods that qualify for the task of near-surface crack detection, special attention is paid to those that allow automatic data acquisition and provide accurate, objective and reproducible results. Common candidate methods are ultrasonic testing (UT), eddy current testing (ET), magnetic flux leakage (MFL) testing, and thermal testing (TT). However, single-method inspection is often ambiguous. This is because each method reacts to changes in specific physical properties of the tested material in the presence of a defect. But, harmless material or geometrical characteristics may also produce similar and often indistinguishable changes in the recorded signals. For instance, MFL is affected by surface roughness and inhomogeneous magnetic properties, and TT's high sensitivity to negligible manufacturing and handling marks and imperfections results in a multitude of false positive indications. These non-defect indications are referred to as *structural noise*, because unlike measurement noise, they are deterministic across repeated measurements. However, structural noise does vary across different inspection techniques, whereas actual flaws are characterized by multi-sensor agreement. This concept lies at the heart of NDT data fusion. On other words, multi-sensor inspection differentiates flaw indications from structural noise, thus providing more reliable assessment.

Sensor fusion can be performed at various levels of signal abstraction [1], each with specific drawbacks and advantages. In particular, decision-level fusion deals with the higher level aggregation of data after individual detection. That is, each signal is first processed individually, and then fed into the fusion algorithm. We argue that decision-level fusion has several advantages over signal-level fusion for NDT inspection. First, unlike signal-level fusion, it does not require obtaining individual signal values at the very same locations by interpolation. Although a spatial association step is still necessary to prepare individual detections for fusion, decision-level fusion can work on the un-interpolated original data. As we will demonstrate here, this approach allows for explicit handling of localization uncertainty, for instance due to registration errors. This is in contrast to signal-level fusion, where accurate registration is crucial [2, 3] because errors can hardly be compensated. Another advantage of high-level fusion for defect detection is the two stage process: the measurements are sifted once for flaws as in traditional single-sensor inspection, and a second time after fusion. This process promises to effectively eliminate many of the false alarms that result from structural noise. A further major benefit is the strategy's modularity. The individual data collection and processing can be carried out independently by the respective experts to tailor the detection process specifically to each inspection method. One practical benefit of the modularity offered by decision-level fusion is that it allows combining individual results, even if fusion was not envisioned in the original testing plan. This also facilitates independence from the type of data source, making it possible to aggregate heterogeneous modalities ranging from manual inspection data to the output of fully automated scanning systems. Consequently, different sources of information can be effortlessly exchanged and the fusion strategy is readily adapted to unknown future changes of input sources.

Although there are recent studies in NDT proposing decision-level fusion, e.g. by weighted averaging [4], hypothesis testing [5], and Bayesian or Dempster-Shafer theory [6, 7], all of these works still rely on image registration and interpolation to perform fusion at common grid points. In fact, we are not aware of any data fusion publication in the field of NDT dealing with *scattered* decision-level fusion, i.e. using the original measurement locations, despite the aforementioned advantages.

We propose a new fusion strategy that combines spatially scattered detection locations to improve the detection performance compared to single-source methods. In this paper, we will address the problem of nondestructive *surface* inspection, although the methodology is quite generic and can be easily extended to the three-dimensional case of volume inspection. Our approach is detailed in section 2 where the general idea is introduced using a simple example. Subsequently, we formalize the problem, describe our method and provide practical techniques for automatic parameter estimation. The developed approach is then applied to experimental multi-sensor NDT measurements by three different inspection methods. Finally, we quantitatively demonstrate the enhanced performance of multi-sensor crack detection using our technique over single-sensor inspection.

## 2  Methodology

Per-sensor detection is done after individual pre-processing, using simple rules such as thresholding. The result of individual defect localization is a set of detections $\boldsymbol{d} = (d_x, d_y)$, also termed *hits*, scattered across the specimen's surface, each exhibiting a certain signal to noise ratio (SNR). For example, consider Figure 1 for an outcome of individual surface inspection using two NDT methods. In cases *a* to *d*, each dot marks a hit according to some detection rule per sensor. Among the hits, there are also false alarms, for instance indicating changes in material properties unrelated to a defect (*structural noise*). Such false alarms are illustrated by cases *b* and *c* in Figure 1. Using single-sensor inspection, these false alarms cannot be distinguished from indications produced by actual defects such as those shown in *a*. A multi-sensor data set, on the other hand, is able to reveal *a* as a real defect by assessing the agreement among different detection methods. Here, agreement is expressed in terms of joint spatial density of hits, taking into account all sensors. Our rationale is that the joint hit density is higher over real defects than in other areas, provided sufficient SNR for at least two sensors. On the contrary, a clear conflict occurs where only one sensor generates hits, and thus the joint density is not significantly increased relative to the individual sensor density. This concept is depicted at the bottom of Figure 1, where for each sensor the spatial density of hits across the specimen surface is symbolized. Only in case *a* both sensors agree in increased spatial density, whereas in cases *b* and *c* the sensors do not agree. Although there is also agreement in case *d* as well, the joint density is not significant enough, indicating the low likelihood of defect presence. This example demonstrates the potential of the joint spatial density as a measure for multi-sensor detection.

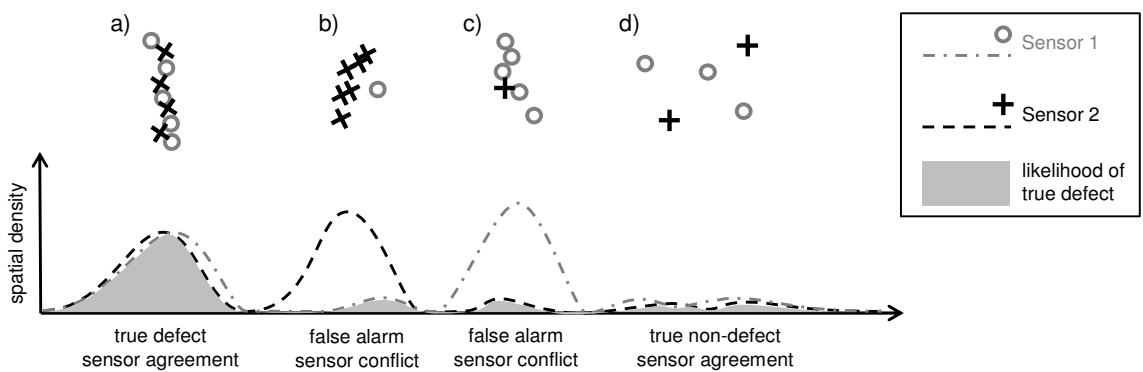

Figure 1: Schematic representation of the principle of the approach. The detection outcome of two different NDT methods are represented by circles and crosses, for four cases a)-d). For each case, the corresponding spatial densitiy per sensor is plotted below. The likelihood of observing a true defect (gray area) depends on both sensors yielding significant hit densities.

We propose evaluating the local hit density as a measure for multi-sensor data fusion at decision level. The first step is detection per individual sensors. The initial detection rules must focus on sensitivity rather than specificity, thus ensuring that all (unknown) real defect indications are retained for the final detection by fusion. At the same time, we would still want to discard as many false alarms as possible. For example, most of the typical signal intensities are near zero. Therefore, a simple thresholding operation significantly reduces the number of potential defect indications per sensor, and thus simplifies the subsequent data fusion efforts. Since the detection method is independent from the proposed fusion approach, the most suitable detection criterion for each sensor can be used.

Moreover, the locations returned by the NDT sensors have to be related, if they are not defined with respect to the same coordinate system. Rather than measuring at the very same locations on the specimen, it is often more practical to mathematically register the individual local coordinate systems after the measurements were carried out. Therefore, for each sensor pair, we find corresponding locations in the sensor data and fit a coordinate transformation model that minimizes the distance between each pair of corresponding locations. Note that, in contrast to fusion at the signal level, this transformation is not used here to interpolate the sensor data values. Rather, for decision-level fusion, the transformations allow to express the locations of the individual detections in any of the sensors' coordinate systems. We note that although each NDT method usually uses gridded measurement positions, the mapped hit locations after registration are not jointly gridded in the common coordinate system, but appear scattered instead.

One central challenge in the decision-level fusion of non-gridded locations is the uncertainty in localization. Two main factors contribute to this uncertainty. First, each sensor's localization ability is limited by the physical resolution as well as the spatial sampling rate. Second, the coordinate transformations computed during spatial registration are inaccurate to some degree, because the manually set point pairs do not perfectly match or the data do not fit the transformation model well enough. To be robust, a fusion approach must adequately cope with the inherent uncertainties about hit positions and must associate nearby points for the purpose of density quantification. This loose concept of "proximity" must therefore be mathematically formalized.

To this end, various non-parametric techniques have been developed such as Mean shift [8], DBSCAN [9], OPTICS [10], Spectral clustering [11], and Kernel density estimation (KDE) [12, 13]. We have selected the framework of kernel density estimation for this study. This choice was motivated by considering that our data space typically has only two or three spatial dimensions, independent from the number of sensors. Therefore, the density can be directly modeled without being affected by the curse of dimensionality. Furthermore, it allows us to evaluate the density at arbitrary positions, not only at the hits. Next, we will formally introduce our method using ideas from KDE.

## 2.1 Kernel density estimation (KDE)

KDE is a nonparametric statistical method to estimate a probability density function $\hat{f}$ from a set of samples $x_i$. The result is a continuous function, computed from a weighted sum of kernel functions $K_h$ with an associated bandwidth $h$, each centered over one of the samples: $\hat{f}(y) = (\sum_i w_i)^{-1} \sum_i w_i K_h(y - x_i)$, with $K_h(x) = \frac{1}{h} K\left(\frac{x}{h}\right)$. Some functions qualifying as a kernel $K$ are the uniform, triangle, Gaussian or Epanechnikov kernel functions. The bandwidth $h$ controls the size of the neighborhood in which each sample influences the density. The choice of the bandwidth is critical for the overall performance of the algorithm. If the bandwidth is chosen too wide, KDE results in an overly smoothed density, thus losing important details of the distribution. On the other hand, if it is chosen too narrow, the estimate adapts too much to the specific realization of the sample set, thus missing the global features of the density. This problem has been well-studied, and several solutions have been proposed [14]. We will describe how to automatically compute a suitable bandwidth for our problem in section 2.3.

The general formulation given above for KDE includes the normalization constant $(\sum_i w_i)^{-1}$, ensuring that the density integrates to one. We proceed with a simpler unnormalized version of KDE by dropping the normalization constant. Furthermore, since the density estimate is a weighted sum with one term per data point, we can partition the data set and aggregate the total density function $\hat{f}$ from the sub-densities $\hat{f}_j$, which include only the samples $x_i$ from each partition $j$:

$$\hat{f}(y) = \sum_j \hat{f}_j(y) = \sum_j \sum_{i \in P_j} w_i K_h(y - x_i)$$

$P_j$ denotes a subset of all points such that partitions do not overlap and the union of all partitions covers the complete data set. This re-arrangement is taken up in the following section to group detections by each sensor.

KDE can be extended to vector-valued samples. To this end, let $\boldsymbol{x_i}$ denote the ith vector-valued (bold face) sample, and let $\boldsymbol{y}$ denote the vector-valued location where to evaluate the density. Multivariate KDE is computed from multivariate kernel functions and an associated bandwidth matrix $H$, which describes the scale and the orientation of the kernels. A special kind of multivariate kernel is the product kernel, defined by $K_H(\boldsymbol{x}) = \prod_j \frac{1}{h_j} K\left(\frac{x_j}{h_j}\right)$, where one univariate kernel for each dimension $j$ is evaluated. The $h_j$ are the entries of the diagonal bandwidth matrix, i.e. product kernels are not arbitrarily oriented in the data space. We use product kernels in our approach, which is described next.

## 2.2  Scattered decision-level fusion

In this section, we develop a new fusion method for our NDT problem based on concepts from KDE. Here, the role of the vector-valued data sample $\boldsymbol{x_i}$ is taken by the two-dimensional hit location $\boldsymbol{d}$ as detected by a single sensor $S_i$ during surface inspection. The fused density for defect detection $\hat{f}$ is computed as:

$$\hat{f}(\boldsymbol{p}) = \left( \sum_{S \in \{S_1, \ldots S_N\}} \hat{f}_S(\boldsymbol{p}) \right) - \max_{S \in \{S_1, \ldots S_N\}} \hat{f}_S(\boldsymbol{p}) \qquad (1)$$

$$\hat{f}_S(\boldsymbol{p}) = \sum_{\boldsymbol{d} \in D(S)} w_d \, K_{\boldsymbol{h}_S}(T_S(\boldsymbol{p}) - \boldsymbol{d}) \qquad (2)$$

$$K_{\boldsymbol{h}}(\boldsymbol{u}) = \begin{cases} \left(1 - \left(\frac{u_x}{h_x}\right)^2\right)\left(1 - \left(\frac{u_y}{h_y}\right)^2\right), & |u_x| \leq h_x \text{ and } |u_y| \leq h_y \\ 0, & else \end{cases} \qquad (3)$$

In Equation (1), $\hat{f}(\boldsymbol{p})$ is the estimated density evaluated at arbitrary location $\boldsymbol{p} = (p_x, p_y)$ on the specimen surface. This density is computed from the sum of all per-sensor densities $\hat{f}_S(\boldsymbol{p})$, each of which only regards detections $\boldsymbol{d} = (d_x, d_y) \in D(S)$ that were found by sensor $S$.

A key element of our method is to explicitly penalize situations in which only a single sensor makes the major contribution to the sum of densities: for each evaluation point $\boldsymbol{p}$, we subtract the maximum single-sensor KDE at this location. This step realizes the quantification of agreement among sensors, because a large fused score now requires at least two sensors with high individual densities. Thus, $\hat{f}$ is expected to behave similarly to the function indicated by the shaded area in Figure 1. Note that in the case of only two sensors, subtracting the maximum is equal to the *minimum* fusion rule, which is a fuzzy *AND*-operator, and is in fact the operation used to generate the shaded area in the figure. However, as more than two sensors are involved, requiring that all sensors indicate a defect seems too strict. Therefore, ignoring the maximum contribution can be viewed as a much milder version of the *AND* fusion rule.

To compute the per-sensor densities $\hat{f}_S(\boldsymbol{p})$, the point of interest $\boldsymbol{p}$ is mapped to the local coordinate system through the coordinate transform $T_S$, where the distance between $\boldsymbol{p}$ and each detection is computed as shown in Equation (2). This measure of proximity is converted to a measure of density by the per-sensor kernel function $K_{\boldsymbol{h}_S}$. Although in principle, the use of any kernel type is valid, we suggest using a compactly supported kernel function like the *Epanechnikov product kernel* as defined in Equation (3). The compact support has the advantage of limiting the spatial influence area of each detection, which is expressed by the kernel bandwidth parameters $\boldsymbol{h} = (h_x, h_y)$, and thus facilitates faster computation than e.g. the non-vanishing Gaussian kernel. We further adapt the kernel function $K_{\boldsymbol{h}}$ so that instead of integrating to unity, its maximum is normalized to one. The rationale behind this choice is discussed in view of automatic bandwidth selection in the following section. In addition, note that instead of transforming the individual detection locations $\boldsymbol{d}$ to the coordinate system of $\boldsymbol{p}$, we follow the converse approach to map the density evaluation locations $\boldsymbol{p}$ to each sensor's system. This simplifies density estimation, because in each local system, the respective kernels are axis-aligned, as will be explained in the next section. This facilitates using a multivariate product kernel, in contrast to general kernels with non-diagonal bandwidth matrix. Figure 2 visualizes the role of the coordinate transformation $T_s$ in our method.

To further adjust the quantification of density, we scale the kernel functions according to the weight $w_d$ per detection. These weights control the influence of each detection on the final KDE. We set each weight proportional to the detection's signal to noise ratio, so that clear indications have more impact on the final density than insignificant ones. Also, the weights offer additional flexibility to regulate the fusion result with regards to specific sensors or different inspection areas.

In total, our fusion approach includes three mechanisms to ensure robustness against false alarms: quantification of density, subtraction of the maximum single-sensor contribution, and decision weighting according to significance.

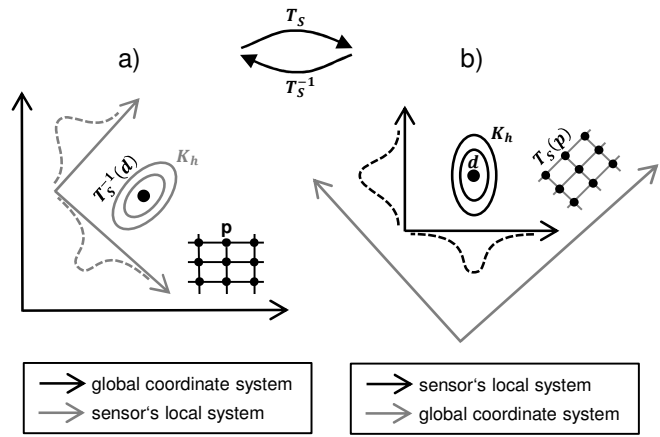

Figure 2: Coordinate transformation during the computation of the fused density. The two black coordinate systems in a) and b) are related through the coordinate transform $T_s$.
a) Coordinate system in which the fused density will be evaluated at gridded points $p$. The coordinate systems of the individual sensors, where the detections $d$ are defined, are not axis-aligned with the global system. In particular, the kernels $K_h$ would require non-diagonal bandwidth matrices.
b) Coordinate system of one of the sensors, given by its measurement grid. For single-sensor density estimation, kernels $K_h$ are axis-aligned to the sensor's system, thus facilitating product kernels. The single-sensor density is then evaluated at the transformed points $T_s(p)$.

## 2.3  Automatic bandwidth estimation

We can use our background knowledge about the nature of our data to deduce the minimum required values for the bandwidth. Intuitively, the density estimator should always be able to smoothly interpolate between neighboring data points. For NDT data, the smallest possible distance between any two detections of the same sensor is given by the known spatial sampling intervals. For a measurement grid per sensor $S$, the two spatial sampling distances are denoted by $\Delta_x^S, \Delta_y^S$ in the sensor's coordinate system. To ensure that neighboring line scans crossing the same defect do not form disconnected density peaks, the kernel functions should at least stretch across one pixel in the sensor image. However, to avoid merging two unrelated indications, the kernels should not be made much larger[1]. Therefore, we propose to use product kernels with minimum bandwidth parameters of $\boldsymbol{h}_s = (h_x, h_y) = (\Delta_x^S, \Delta_y^S)$ for each sensor. It is natural to use product kernels for gridded individual measurements, because the bandwidths directly correspond to the physical pixel dimensions. Note that, in conventional KDE, broader kernels are necessarily flatter to maintain unit integral. We, in contrast, fix the kernel's peak value to one, independently from its spatial extent. By following this procedure, each sensor contributes equally to the joint density, regardless of its spatial localization characteristics. Figure 3 illustrates this concept.

The aforementioned kernel size is a minimal setting. In practice, the most significant factor contributing to the localization uncertainty may not be the spatial sampling, but inevitable registration errors. The kernel sizes should be set large enough to smooth out these unwanted variations and to associate poorly registered indications. As a general approach, we propose the following kernel size $\boldsymbol{h}_s = (h_x, h_y) = (\max\{\Delta_x^S, \hat{u}\}, \max\{\Delta_y^S, \hat{u}\})$, where $\hat{u}$ denotes an estimate of the localization uncertainty, for instance the mean registration error. However, note that in this approach, the advantage of having spatially accurate sensors may be lost. Also, with increasing kernel size, closely situated defects become harder to separate. Therefore, our density-based approach will substantially benefit from high-quality registration by facilitating small kernel sizes.

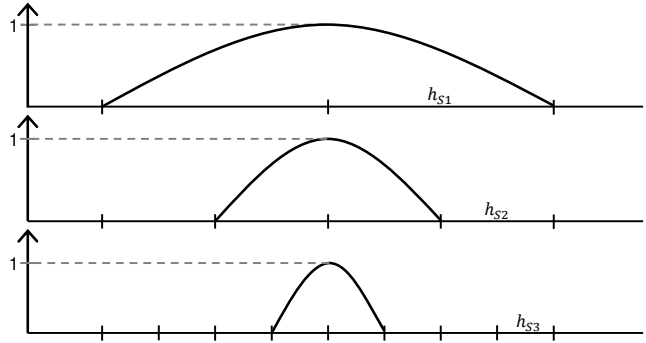

Figure 3: Kernel functions with automatically computed bandwidths, for three sensors with different spatial sampling distances, e.g. in mm. For clarity, only one spatial dimension is shown per sensor. Registration errors also influence the kernel widths, but this is ignored in the depiction for clarity. Moreover, if data weights were regarded, the kernel amplitudes would be scaled.

---

[1] In fact, some NDT methods might have coarse spatial localization ability due to physical limitations, regardless of the spatial sampling rate. Those sensors are characterized by spatially slowly varying signals. Because in such cases, the smallest possible distance between any two detections is larger than one pixel, it is sensible to increase the kernel size.

# 3 Application to experimental data

To demonstrate the fusion technique's performance under realistic conditions, a test specimen containing 15 surface flaws was inspected using three different NDT methods. In this section, we describe the specimen, the individual data collection and processing as well as the application of our fusion algorithm. Finally, we quantitatively evaluate the effect of various conditions on the fusion result, and compare fused detection against single-sensor detection.

## 3.1 Specimen

The ring-shaped test specimen is a bearing shell [15, pp. 173-175] made of surface-hardened steel. As illustrated in Figure 4, it has an outer diameter of 215 mm and is 73 mm long in its axial direction. Cracks are simulated by 15 machined grooves that are regularly spaced across the surface. These grooves were created by electrical discharge machining and vary in depth from 11 to 354 µm, as detailed in Table 1. Grooves have constant lengths of 1 mm and their openings vary between 25 µm and 51 µm. The specimen's surface is uncoated and its roughness is very low, thus enabling high-quality near-contact measurement.

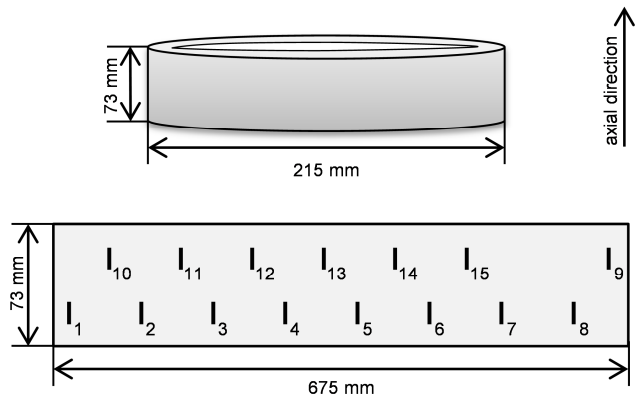

Figure 4: Schematic view of the ring specimen, not to scale. Top: outer proportions of the ring. Bottom: unrolled outer surface. Short vertical lines indicate the positions of the 15 grooves.

| Groove | 1 | 2 | 3 | 4 | 5 | 6 | 7 | 8 | 9 | 10 | 11 | 12 | 13 | 14 | 15 |
|---|---|---|---|---|---|---|---|---|---|---|---|---|---|---|---|
| Depth / µm | 354 | 224 | 170 | 105 | 82 | 61 | 57 | 53 | 43 | 40 | 39 | 27 | 29 | 20 | 11 |

Table 1: Groove depths. Labels correspond to those shown in Figure 4.

## 3.2 Individual measurements and processing

For crack detection, thee different inspection techniques were used that are well-suited for the automated nondestructive evaluation of cracks in ferromagnetic metals. The first method is called *eddy current testing* (ET). An excitation coil is run over the specimen surface. Using this coil, an alternating current induces circular eddy currents in the specimen's near-surface region. These currents are blocked by the presence of defects, thus affecting the impedance of the probe coil, which is the measured signal. The second method employed here is the laser-induced *active thermography testing* (TT). A high-power laser is run across the specimen to locally heat up the surface. In defect-free regions, the heat flow is able to dissipate, whereas defects cause localized heat accumulation. An infrared camera monitors the heat flow decay and generates a digital image sequence for processing. The third method is *magnetic flux leakage testing* (MFL). The specimen is exposed locally to a static magnetic field, which spreads inside the ferromagnetic material. At near-surface defects, the field is forced to "leak" out of the specimen into the air, although air has lower magnetic permeability than the specimen. This leakage can be detected by magnetic field sensors, such as *giant magneto resistance (GMR)* sensors. The following three inspections were performed sequentially during the course of about one year.

ET was carried out at an excitation frequency of 500 kHz. An automated scanning device rotates the specimen under the fixed probe. Signal processing is based only on the imaginary part of the measured impedance. The obtained one-dimensional signals are preprocessed by high-pass filtering for trend correction, and by low-pass filtering to improve SNR. An image is formed by stacking the line scans in axial direction of the ring.

The MFL data were collected using the same scanner as for ET, and a GMR sensor array developed at BAM [16]. Using these gradiometers, the normal component of the magnetic stray field was measured while the specimen was locally magnetized. Preprocessing comprised trend correction by high-pass filtering per line scan, and an adaptive 2D wiener filter[2] for noise suppression. The image was then Sobel-filtered to highlight the steep flanks that are generated by the gradiometers over the grooves.

Thermography testing was performed by rotating the specimen under a 40 W powered laser while recording with an infrared camera. The movie frames were then composed to form an image of the specimen surface. This image is processed by 2D background subtraction using median filtering, and noise was suppressed by an adaptive 2D wiener filter.

We note that the presented signal acquisition is tailored to the known groove orientation. In a realistic setting, we would have to perform a second scan for ET and MFL testing to detect any circumferentially oriented defects as well. We further emphasize that for our specific ring specimen, the GMR sensors yield far superior results compared to ET and TT, and would suffice by themselves for surface crack detection. However, such performance is not guaranteed for other materials or in case of suboptimal surface conditions, so that a multi-method approach is still in demand. To take advantage of as many independent sources of information as possible during fusion, we intentionally lowered the quality of the MFL image before preprocessing and detection. This was done by separating the true defect indications from the background signal variations using the shift-invariant wavelet transform [17], and by reconstructing the signal with a factor of 0.1 for the noise-free component. Although this does not simulate a lower-quality MFL measurement in a physically realistic way, the approach allows making use of the acquired GMR signals as a third source of information. We justify this alteration in favor of demonstrating the capabilities of our fusion technique in other settings where individual inspection is in fact not reliable enough. Therefore, for the rest of the article, we consider only this modified version of the MFL data.

---

[2] We used MATLAB's function *wiener2*. See http://de.mathworks.com/help/images/ref/wiener2.html

To convey an impression of the signals, we show an exemplary portion of each preprocessed inspection image in Figure 5. The displayed part of the specimen surface is a 10 mm by 6.5 mm region around groove nr. 13 which is quite shallow, and thus generates relatively weak indications. The figure demonstrates the different signal patterns among sensors, concerning both the groove and the background variations. Also, the different pixel sizes are evident. A related plot is shown in Figure 6, where one-dimensional line scans crossing the groove reveal more clearly the individual sensor responses. The different spatial sampling positions are demonstrated by the line markers. Table 2 offers a quantitative comparison of the individual data sets.

After individual preprocessing, the same detection routine was used for all three images to extract hit locations and confidences, which will later be fused at the decision level. To this end, we convert the signal intensities to confidence values and apply a threshold to extract only significant indications. Confidence values are computed by estimating the distribution function of background signal intensities from a defect-free area. This estimate serves as the null distribution $P_{null}(x)$ in the significance test. For each image pixel's intensity $y$, the probability $P_{null}(y \geq x)$ is computed as the confidence. Pixels are considered significant here, if their confidence exceeds 99 %. Additionally, only those hits that are local maxima with regard to their neighboring pixels along the horizontal axis (thus crossing the grooves) are retained. This constraint further filters many false alarms while making the detection results invariant to different peak widths. After registration to a common coordinate system based on manually defined location correspondences in the data, the final set of hits from all sensors is plotted in Figure 7. Obviously, the false alarms considerably outnumber the actual groove detections. This is due to our sensitive detection rules, intending that no actual defect is missed during individual processing. Of course, in a single-sensor inspection task, a much more stringent detection criterion is appropriate to limit the number of false alarms. However, this possibly leads to worse sensitivity to small material flaws. In contrast, our data fusion approach is supposed to discard most false hits while maintaining high sensitivity to small defects.

We will now briefly compare the individual sensor results. In contrast to ET and TT, the MFL hits cluster spatially. This is because the background variations in this data set are not homogeneous, possibly due to inhomogeneities in the internal magnetic field. MFL data are missing in the strip between the two groove rows. Interestingly, in Figure 7, spatial defect-like patterns are formed, although the specimen is not expected to contain any flaws other than the known grooves. For example, see the vertical lines from TT

(e.g. indicated by the right arrow), or the diagonally oriented lines from ET indicated by the left arrow. As previously discussed, using individual inspection, it is not easy to classify these obvious indications as structural indications or flaws. In spite of their regular structure, we label these patterns as non-defect indications during the following evaluation, if our multi-sensor data set is not able to give a reliable confirmation. On the other hand, there are a few off-groove locations where different sensors behave consistently. These regions could in fact represent unknown but real defects, and will therefore be excluded from the following evaluation. Note that detections within disregarded areas are not shown in this figure. Moreover, the confidence associated with each detection is not shown, because all hits exceed the chosen threshold of 99 % as explained before.

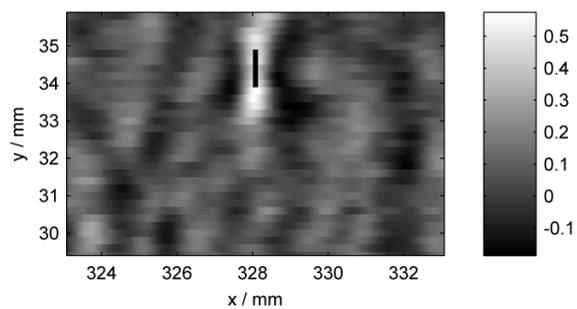

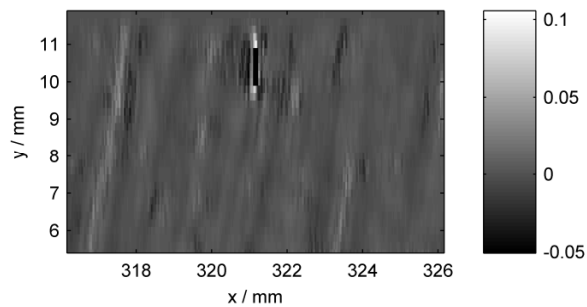

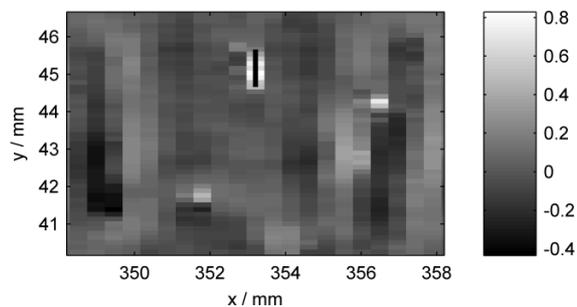

Figure 5: Preprocessed sensor intensity images, zoomed to a region around the groove nr. 13. Higher intensities correspond to indications. Top: ET, Middle: MFL, Bottom: TT. The vertical black line marks the location of the groove. Each image is shown in the respective sensor's coordinate system, thus explaining the different axis labels.

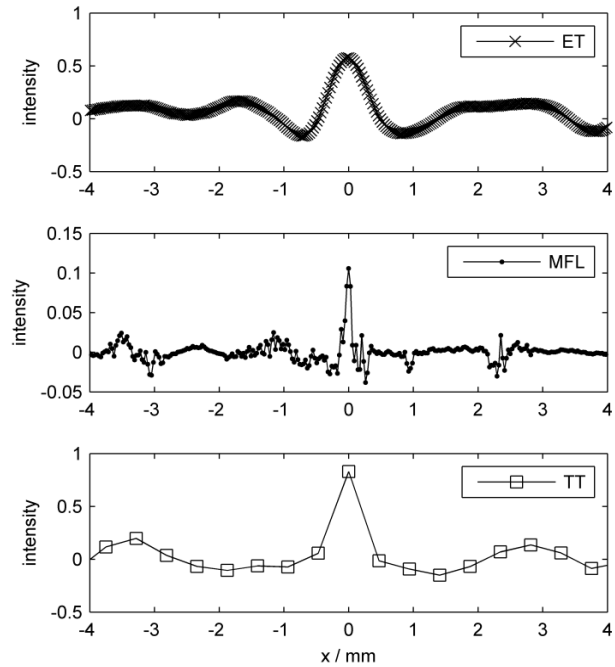

Figure 6: Preprocessed line scan per inspection method around groove nr. 13. The signals are shifted so that each peak value is located at $x = 0$. Note the different y scales.

Table 2: Quantitative properties of the individual data sets

|  | ET | MFL | TT |
|---|---|---|---|
| $\Delta_x$ in mm | 0.029 | 0.029 | 0.469 |
| $\Delta_y$ in mm | 0.200 | 0.200 | 0.126 |
| Width of a typical indication, in mm | 2≈69$\Delta_x$ | 0.6≈20$\Delta_x$ | 0.5≈$\Delta_x$ |
| Avg. nr. of hits per pixel | 0.0023 | 0.0031 | 0.0068 |

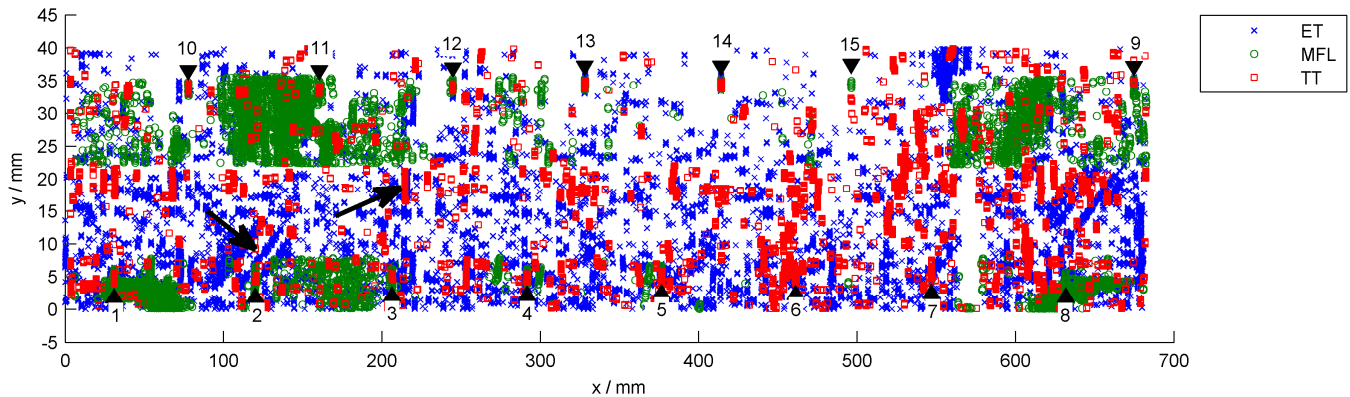

Figure 7: Hit locations per sensor in a common coordinate system similar to that in Figure 4. Axes are not to scale. The tips of the triangle markers indicate the groove positions. The arrows indicate crack-shaped false positives from ET (left arrow) and TT (right arrow). A colored version of this figure appears in the article online.

## 3.3 Fusion and final detection

To compute the kernel density per sensor, we used Alexander Ihler's KDE Toolbox for MATLAB [18]. The fused density, according to Equations (1)-(3), is a continuous function that must be evaluated at discrete locations. In fact, to circumvent the discrete sampling, a multivariate mode-seeking algorithm could be used for detection. However, for simplicity, we set up a discrete evaluation grid that is designed fine enough to not miss any mode of the density. Modes are then traced similarly to per-sensor detection by finding local density maxima along parallel lines on the specimen surface. Figure 8 displays the principle of the detection. Finding local maxima along one-dimensional lines is straightforward due to the density's smoothness, and also makes the detection results more stable across different kernel sizes.

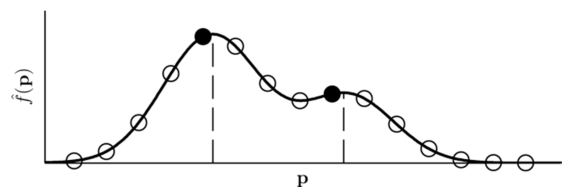

Figure 8: Final detection by evaluating the fused continuous density function at gridded points (circles) and finding local maxima (filled circles) among them. These detections approximate the true density modes (dashed lines).

The final detections after fusion are presented in Figure 9. Most of the single-sensor false alarms from Figure 7 were discarded by our fusion method by recognizing the sensor conflicts. Yet, there are a considerable number of remaining false hits. These spurious detections originate from single-sensor hits that overlap purely by chance. Nevertheless, all grooves but the shallowest one, nr. 15, clearly stand out against the false alarms considering the fused density measure, which is represented by the marker colors in Figure 9. Because higher values of the fused measure correspond to increased defect likelihood, a threshold can be applied to produce a binary decision. In the following, the detection performance will be quantitatively assessed under various conditions.

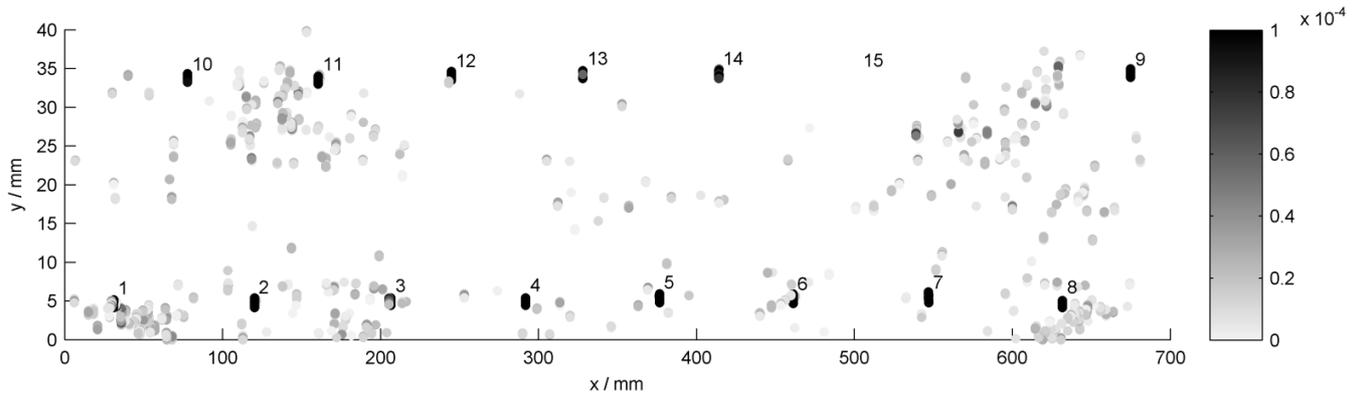

Figure 9: Result of decision level fusion. Darker markers correspond to increased detection confidence. The colors range from zero (white) to the fused intensity at the shallow defect nr. 14 or higher (black). Axes are not to scale.

## 3.4 Evaluation

In the following sub-sections, our fusion method is quantitatively evaluated with regard to the presented specimen. This evaluation focuses on detectability, meaning the ability to distinguish between grooves and background in the fusion result. Consequently, the ability to accurately localize a defect after fusion is not a part of this evaluation. The evaluation is realized by manually labeling detections near the known groove locations as "on-groove", and others as "off-groove". Specifically, for each method to be evaluated and for each groove, we define a polygon, inside which hit locations are labeled as "on-groove". In the same way, several regions on the specimen surface are marked to be excluded from the evaluation. These are areas near the border of the specimen, indications that result from experimental modification of the specimen surface, off-groove areas where real unplanned defects exist (which would otherwise be counted as false alarms), and a margin around each groove polygon. Excluding this margin is necessary, because the defect indications might exceed the actual defect region on the specimen, as is seen in Figure 5. Hits in this margin usually have reduced intensity compared to the center of the flaw. Therefore, if they were included in the on-groove class, these intensities might not exceed the detection threshold and thus the groove would often seem to be detected incompletely. On the other hand, if they were counted as off-groove, the number of false alarms would be spuriously increased, because the intensities are often still larger than the background signal. Not only are all of these disregarded regions removed from evaluation after fusion, but already the hits in these regions are excluded from the density estimation, so that they don't affect the density in the surrounding regions. Furthermore, to evaluate detection performance per flaw depth, after fusion each groove is assessed individually while ignoring all others.

We note that there are several issues concerning the comparability of different detection methods using this evaluation approach. The fundamental issue is that detection, and therefore evaluation, is not carried out per (hit) region, but per point. This approach complicates evaluation, because different detection methods (e.g. single sensor detection and fused detection) yield different numbers of hits at different locations. This heterogeneity is compensated here by tailoring the ground truth to each investigated detection method, using the described polygons. Yet, results are never fully comparable, because each is evaluated on a slightly different basis. Nevertheless, the evaluation was designed to be as fair and expressive as possible.

Detection performance is measured by area under the precision-recall curve. This method is similar to the well-known area under the Receiver Operating Characteristic (ROC) curve, but we compute precision *prec = #true hits/#all hits* instead of false alarm rate *FAR = #false hits/max possible number of false hits* and plot it against the true positive rate *TPR = #true hits/#max possible number of true hits*. Precision is suited for class-imbalanced problems. This is clearly the case here, because the groove regions are much smaller than the background region. Consequently, $FAR$ is always small, even if there are many times more false alarms than true hits. In contrast, $prec$ relates hits to hits, rather than hits to non-hits. Furthermore, we do not compute the area under the whole curve, but only for the curve region where recall = true positive rate = $\#true\ hits/\text{max possible number of true hits} > 0.5$. We denote this measure by *AUC-PR-0.5*. This focuses our evaluation on thresholds that are low enough to ensure that at least half of a groove is detected. Furthermore, a single false alarm hit with higher intensity than the groove suffices to force the curve down to zero precision for small true positive rates, i.e. high thresholds, and therefore dominates the whole AUC measure. This is another reason for ignoring the left half of the diagram in the computation of AUC-PR-0.5.

Using this evaluation framework, we investigate the performance of our fusion approach in the following.

## 3.4.1 Influence of kernel size

In section 2.3, we presented an automatic method to set up the kernel bandwidth parameters. In this section, we quantitatively compare the influence of different kernel sizes on detection performance in our data set.

For our inspections, the localization uncertainty due to registration errors[3] is up to $\hat{u} = 0.2$ mm. We take this value as the basis for the range of suitable kernel sizes. Specifically, the following kernel sizes are investigated: $\boldsymbol{h}_s = (h_x, h_y) = \left(\max\{\Delta_x^S, a\hat{u}\}, \max\{\Delta_y^S, a\hat{u}\}\right)$ with $a \in \{0, 0.5, 1, 2, 5, 10\}$. See Table 2 for the specific values of $\Delta^S$. For $a = 1$, we obtain the setting proposed in section 2.3. For $a = 0$, the kernel sizes are computed only from the sampling distances of each sensor, disregarding the registration errors. For high values of $a$, the effect of over-smoothing should be observable. For all investigated settings, the density was evaluated at the same locations. These locations are defined by a super-sampled measurement grid of one of the individual sensors.

---

[3] As measured by the following: One location on the specimen surface corresponds to point A in the coordinate system C1, and to point B in the coordinate system C2. The Euclidean distance between point A mapped to C2 and point B is a registration error. From the distribution of these errors a suitable summary statistic is generated.

Figure 10 compares the results. The detection performance appears to be stable across a broad range of kernel sizes. Perfect detection (AUC-PR-0.5=0.5) is possible for almost all defects. However, at the shallowest groove nr. 15, SNR in the individual sensors is not sufficient to yield multi-sensor hits. In contrast, the second-shallowest groove nr. 14 is detectable, although performance depends on the kernel size. The SNR at this region is relatively low, so that larger kernels increasingly produce false indications in the fused image. However, the smallest investigated kernel size is also not suited. For this groove, the optimal kernel size indeed matches the registration error ($a = 1$). The deeper grooves, indicated by lower indices, are readily distinguished from the large number of per-sensor false hits. Only for $a = 0$, the kernels apparently are too narrow to associate the individual hits that indicate a groove. This is due to our relatively large registration error, which this setting does not account for. Interestingly, no over-smoothing effect due to large values of $a$ can be observed here. This could be attributed to the other implemented mechanisms for false alarm rejection, such as the SNR-based weights $w_d$ and the subtraction of the maximum individual sub-KDE. At very large sizes $a = 10$, there are two outlier flaws in the evaluation. The seemingly poor performance is due to a spatial shift of the fused indications, thus exceeding groove region that was pre-defined for the evaluation.

We suggest using $a = 1$, that is setting the kernel sizes for sensor S to
$\boldsymbol{h}_s = (h_x, h_y) = (\max\{\Delta_x^S, \hat{u}\}, \max\{\Delta_y^S, \hat{u}\})$, if information about localization uncertainty $\hat{u}$ is available.

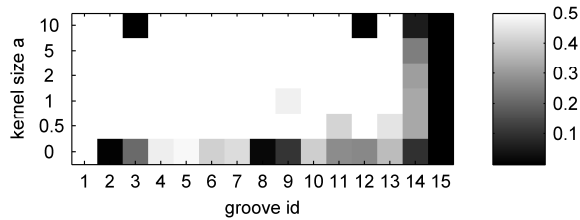

Figure 10: Evaluation of different kernel sizes for each groove, measured by AUC-PR-0.5. Larger is better, maximum possible score is 0.5. Grooves with higher indices are shallower and thus harder to detect.

### 3.4.2 Influence of weights

In the previous experiment, the individual sensors' detections were weighted by factors $w_d$ to take into account the local SNR. To this end, SNR is computed from a null distribution of background signal intensities, similarly to what was described in section 3.2. But in instead of forming probabilities, we compute a z-score $w_d = (x - \hat{\mu}_{null})/\hat{\sigma}_{null}$ from the respective signal intensity $x$ under the null distribution using robust estimates of $\mu_{noise}$ and $\sigma_{noise}$. This linear transform normalizes the intensities among different sensors while preserving the intensity information.

In this section, we repeat the experiment of section 3.4.1, but without weighting the individual hits by setting all weights $w_d = 1$. Therefore, the fused density is only influenced by the proximity of the individual hits. This setting represents inspection results for which no measure of confidence is available.

See Figure 11 for the resulting detection performance, based on unweighted hits. In comparison to the weighted results shown in Figure 10, detection performance is generally worse. In particular, groove nr 14 is barely detectable with this approach. However, for larger kernel sizes at $a = 5$, optimal detection results are still possible for most defects. We explain this by the observation that $5\hat{u}$ equals the groove length for our specimen. Therefore, all hits from a groove contribute to the fused intensity, whereas for smaller kernel sizes, different parts of the grooves are unrelated. This is clearly seen at $a = 0$, where performance is the poorest, in agreement with the previous experiment. Note that the general trend of improved performance for larger kernels can partly be explained by a bias in the evaluation regime: Larger kernels lead to smoother densities that form fewer local maxima ("peaks"). Therefore, the number of potential false alarms after fusion decreases, which benefits precision.

We conclude that, if possible, the weighted approach should be favored over the unweighted method. If weighting is not possible, kernel sizes should not be made too small to fully exploit the density information. We further suggest setting the kernel size smaller than or equal to the expected flaw length.

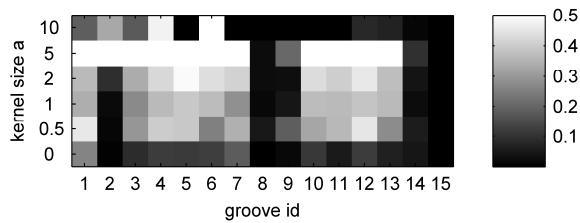

Figure 11: Evaluation of different kernel sizes, for unweighted hits.

### 3.4.3 Influence of individual sensors

The choice which sensors to fuse may have an impact on the detection capability. In particular, we are interested in the contribution that each individual source has on the fusion result. Therefore, we compute fused densities while excluding one of the three inspections, and compare these results to fusion using all three data sets. Based on the results of the previous two experiments, we choose a weighted KDE technique with kernel sizes set according to $a = 0.5$ which proved to perform well. The same evaluation locations were used as in section 3.4.1. We would like to remind the reader that our fusion approach involves suppressing single-sensor detections, i.e. areas in which the joint density is generated by only a single sensor. Therefore, if one sensor misses a defect due to poor SNR, both other sensors must yield hits to indicate this defect in the fusion result. In our experiment, the exclusion of one of these sensors will lead to strongly decreased performance.

As Figure 12 shows, the full set of inspections is necessary to achieve the best detection performance here. Each two-sensor subset of inspection methods shows slightly different effects. Apparently, the thermographic data mainly help in detecting the shallow grooves 10-14. However, the same inspection seems to have missed the flaws 2, 3, 8 and 9, because the information from both MFL and ET is crucial for detection here. Among the deeper grooves, nrs. 1, 4, 5, 6 and 7 are perfectly found using any two-sensor configuration, thus indicating that they are clearly represented in all three measurements. For the shallower set of grooves, the inclusion of the ET inspection seems to bring slightly smaller gains than including the MFL or TT inspection. In particular, the second shallowest groove, nr. 14, is readily detected using a combination of only MFL and TT. This is especially interesting in light of the following experiment, where single-sensor performance is contrasted with fusion performance.

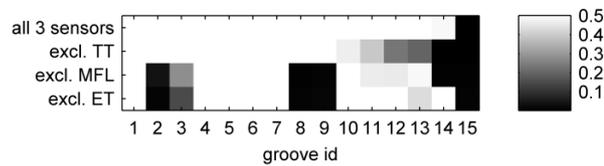

Figure 12: AUC-PR-0.5 values for fusion using different sub-sets of inspections.

### 3.4.4 Comparison against single-sensor detection

Here we quantify the gain of fusion over single-sensor inspection. Again, we choose a weighted KDE technique with kernel sizes set to half the registration error ($a = 0.5$). Unlike the previous experiments, which were all conducted in the same spatial coordinate system, here each single sensor has its own system. Since we compare each individual sensor against the fusion result, it is natural to evaluate the density from each of these individual comparisons in the respective single sensor's coordinate system. We super-sample each measurement grid to avoid missing density peaks. Note that in all three comparisons, the same density function is evaluated, albeit at different locations. To simplify the evaluation, we report only a single fusion result by selecting the worst outcome per groove among the three results.

For fair comparison of single-sensor inspection against fusion, only the pixels that represent a local intensity maximum ("peak") define the set of evaluated single-sensor locations. In particular, these points are not chosen according to their intensity. This set of hits is further filtered by the regions that should be ignored throughout the evaluation (see section 3.3). A performance curve is formed by varying a threshold over this distribution of intensity peak values and counting precision and recall for each threshold value. In contrast, for the assessment decision-level fusion, the per-sensor detections are first judged according to their confidence, as described in section 3.2. From these hits, our fused density is then computed, which is in turn subjected to local maximum detection. The density values at these peak positions are then used to form the ROC curve for the fusion method.

At this point, we emphasize that the results presented in this article are not representative for the general performance of each individual inspection method. It is possible that better individual results than shown here may be obtained by optimizing e.g. the specimen preparation, the sensors or the processing routines. Rather, in the following our focus is to demonstrate that our technique is able to perform well in the face of imperfect sources of information. Moreover, the single-sensor results are not fully comparable, because the inspected area of the specimen differs between sensors. Instead, each single-sensor inspection should only be compared against the fusion result.

We report AUC-PR-0.5 values in Figure 13. According to our evaluation, the presented fusion technique achieves perfect detection results of AUC-PR-0.5 = 0.5 for the first 12 flaws and outperforms single-source detection for grooves not deeper than 61 µm (groove nr. 6). Interestingly, although the performance index drops below 0.05 for each individual testing technique from groove nr. 10 onwards, high performance above 0.4 is still possible using our fusion approach. This means that most actual defect indications along a flaw generate indications in the fused image that dominate the many off-groove false alarms, which can therefore be suppressed. In contrast, although the individual groove indications may surmount the measurement noise, they are not significantly different from the many off-groove indications, thus impairing precision. The last and shallowest groove (11 µm) is not reflected in the fused density image, because only the MFL sensor is able to generate hits here. However, at the two shallow defects nr. 13 and 14, fusion performance is still above 0.28. Because we report the worst out of three density evaluations here and also the ground truths differ between experiments, this value is lower than the result presented in the previous section, where already a two-sensor combination provided perfect detection. In contrast, no individual sensor is able to indicate these shallow grooves at acceptable levels of precision.

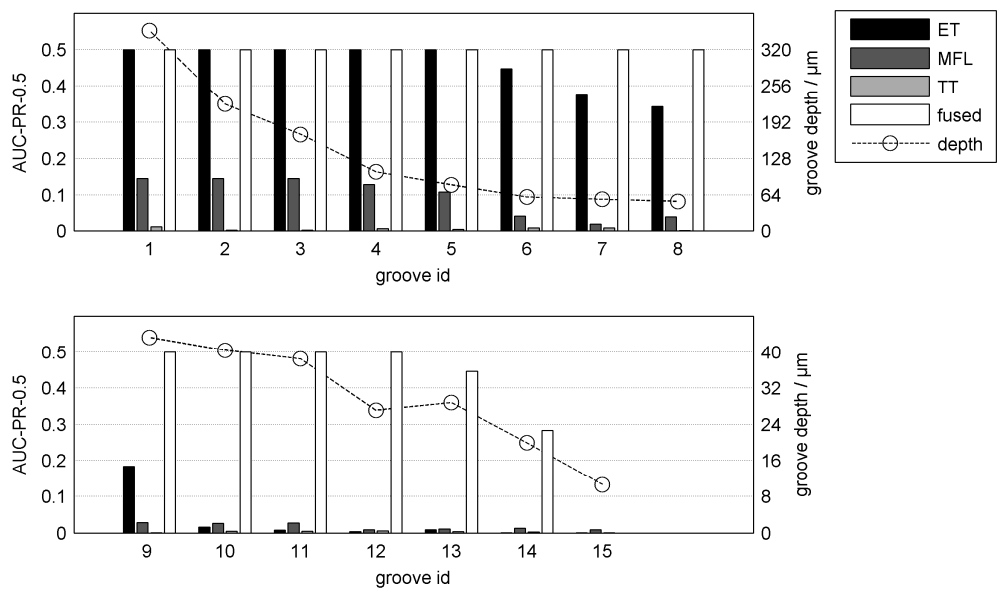

Figure 13: Per-groove performance comparison (bars) of single sensor inspection versus the fusion approach. The maximum possible score is 0.5 (left vertical axis). Here, "fused" actually denotes the worst per groove out of the three

density evaluations (see section 3.4.3). The set of grooves is divided into two sub-figures for clarity. Groove depth is indicated by the plotted line corresponding to the right vertical axis. Note the different axis scales for groove depth in the two subplots.

# 4 Discussion

Our experiments demonstrate that our density-based approach is well-suited to incorporate indications from heterogeneous sensors. The number of false alarms can be strongly reduced relative to single-sensor inspection while retaining most of the defects. Regarding our evaluation index, for all grooves but the shallowest one, the fusion method performs as well as or better than single-sensor detection. The performance gain is most pronounced for the shallower defects, which usually generate less significant indications.

We note that the principle to quantify agreement among sensors requires that all sensors yield redundant information about the object of interest, e.g. near-surface cracks in our case. That is, in NDT, all sensors must respond to the same flaw type in the same size range. If, in contrast, one of the sensors reports a defect that is not detectable by the other methods, it will be discarded as a "false alarm" by our technique, since it is not designed to fuse complementary information.

Another point concerns the relationship between fusion performance and spatial uncertainty. Specifically, the fusion performance is expected to improve with registration accuracy. This is because kernels can be made narrower for smaller registration errors, and therefore the likelihood of a non-defect-related indication due to spurious multi-sensor agreement is reduced. In any case, the actual registration error must be quantified to set the kernel size accordingly. Moreover, the fusion technique strongly benefits from realistic estimates of the local signal to noise ratios, which enter the fused density through weights. For the final detection after fusion, a threshold could be chosen to retain only the significant density peaks. Thus, only few parameters (localization uncertainty, density threshold) fully describe the methodology and are usually readily determined. Furthermore, if one is unsure about a fused indication, the original individual detections can always be reconsidered to collect additional evidence for or against the presence of a defect. After all, the density-based approach spatially associates neighboring hits and thus may serve as the basis for multi-sensor detection after feature extraction. For example, our density measure identifies narrow regions of increased defect likelihood. These regions can further be assessed by extracting features from each individual sensor in this region, which could be combined by some classification algorithm to reach a final conclusion. Concerning the number of sensors, we suggest using at least three different sources of information, as presented here. However, further investigations show that improved performance over single-sensor inspection is possible already for two sensors. Note that the more fusion inputs are provided, the higher the likelihood of the purely coincidental agreement between at least two sensors. Therefore, our proposal to subtract the largest contribution from the sum of individual densities will have to be extended if even more sensors are included.

We would also like to point out the limitations of the current study. Whereas the eroded grooves facilitate detection assessment for well-defined defect depths, their linear shapes do not resemble natural defects. Also, whereas the orientation of our flaws is well-defined, natural defects often vary in orientation. Therefore, directionally sensitive measurements, such as ET using a differential probe and MFL using gradiometers, must be carried out multiple times in different directions. However, because this issue is only relevant for per-sensor detection prior to fusion, it is not further elaborated here.

## 5   Conclusion / Outlook

We developed a density-based method for the fusion of spatially scattered data and applied it to sensor signals from the nondestructive testing of a bearing shell. This high-level fusion approach has the advantage of being independent from the processes that generate the scattered points. Three different mechanisms are implemented to increase robustness against false alarms. Practical suggestions on how to determine the parameters, especially the kernel size, are given. We quantitatively evaluated our technique using a defect detection experiment. The results demonstrate that single-sensor inspections of the specimen are outperformed by the proposed technique, especially for defects that are too shallow to be reliably indicated otherwise. Moreover, the proposed method is quite generic, as it receives spatial locations from single-source detection routines and returns areas of multi-sensor agreement. Therefore, it may be applied for detection tasks in other domains, such as medical image fusion.

## Acknowledgements

We acknowledge the help of our colleagues in divisions 8.4 and 8.7 for kindly providing the measurements and for their support: M. Pelkner, R. Stegemann, T. Erthner, R. Pohl, M. Kreutzbruck, M. Ziegler, D. Mikolai and C. Maierhofer.